\documentclass[10pt,a4paper]{article}
\usepackage[utf8]{inputenc}
\usepackage{array}
\usepackage{graphicx}
\usepackage{amsmath}
\usepackage{amsfonts}
\usepackage{amssymb}
\usepackage{caption}
\usepackage{subcaption}
\usepackage{float}
\usepackage{multirow}
\usepackage{rotating}
\usepackage{cite}
\usepackage[colorlinks=true, allcolors=blue]{hyperref}
\makeatletter
\newcommand\extralabel[2]{{\edef\@currentlabel{\@currentlabel#2}\label{#1}}}
\makeatother 

\begin{document}
\author{Sagardeep Talukdar\textsuperscript{a}, Riki Dutta\textsuperscript{a}, Gautam Kumar Saharia\textsuperscript{a} 
	and \\ Sudipta Nandy \textsuperscript{a}\thanks{email: sudiptanandy@cottonuniversity.ac.in}\\
\\
\textsuperscript{a}Department of Physics, Cotton University, Guwahati 781001, India}

\title{Linear interference and systematic soliton shape modulation by engineering plane wave background and soliton parameters}

\maketitle	

\begin{abstract}
We investigate linear interference  of a plane wave with different localised waves using  coupled Fokas-Lenells equation(FLE) with four wave mixing (FWM) term. We obtain localised wave solution of the coupled FLE by linear superposition of two distinctly independent wave solutions namely plane wave and one soliton solution \& plane wave and two soliton solution. We obtain several nonlinear profiles depending on the relative phase induced by soliton parameters. We analyse the linear interference profile under four different conditions on the spatial and temporal phase coefficients of interfering waves. We further investigate the interaction of two soliton solution and a plane wave. In this case we notice that asymptotically, two solitons profile may be similar or different from each other depending on the choices of soliton parameters in the two cases. The results obtained by us might be useful for applications in soliton
control, a fiber amplifier, all optical switching, and optical computing.
Further we believe that the present investigation would be useful to study the linear interference pattern of other localised waves of FLE.\\

\textbf{\textit{Keywords:}} Fokas-Lenells equation, Four Wave Mixing, Soliton, Linear interference
\end{abstract}

\vspace{5mm}
\section{Introduction}
It is well known that the nonlinear Schr\"odinger type equations namely, nonlinear schr\"odinger  equation (NLSE)\cite{shabat}, derivative NLSE (DNLSE)\cite{kaup}, higher order NLSE (HNLSE)\cite{hirota1}, Fokas-Lenells equation(FLE) can describe the propagation of ultrashort pulses through an optical fiber.  In comparison to other equations, the significance of FLE is due to the presence  of the spatio-temporal dispersion term.   
 FLE was first shown to be integrable by Fokas and Lenells in their original paper\cite{mainpaper}. 
Bright and dark soliton solutions of FLE are obtained using a handful of methods namely,  Hirota bilinear method \cite{Matsun1, Matsun2, sagar, riki}, neural network method\cite{saharia} and by Backlund transformation \cite{vekslerchik}. Breather  solution   \cite{ruomeng}, rogue wave solutions \cite{chen,xu}, algebro-geometric solution  \cite{peng} are among the other reported results.\\
FWM is an important nonlinear physical  phenomena having significance in both nonlinear optics and Bose Einstein condensation (BEC).  In nonlinear optics, FWM is caused by the dependence of the refractive index on the intensity of optical fiber but it critically depends on the channel spacing and fiber chromatic dispersion. In BEC, FWM is defined as a pair transition term. The study of FWM in BEC is important because it can bring in an abundance of nonlinear matter wave structures with different backgrounds.\\
FWM in nonlinear optics and in BEC has been studied earlier
with coupled NLSE \cite{priya,lu, ling1,ling2,ling3,qin yan,pan,rao}, and coupled Hirota model \cite{han}. \\
The study on multi-component FLE specially in the presence of FWM however, has not been done earlier. It is worthwhile to mention that the study of coupled optical fields is specially important for the Wavelength Division Multiplexing(WDM) technology where channel spacing plays an important role to increase bandwidth.\\
 It is interesting to note that in presence of FWM term in the nonlinear Schr\"odinger equation the linear superposition of solutions is possible which is uncharacteristic to nonlinear differential systems. Untill now studies have been carried out for  NLSE and higher order NLSE with FWM term.  So it is natural to ask whether or not the linear superposition of solutions exists for coupled FLE with FWM term. However, to the best of our knowledge the study on the linear interference between two component FLE systems in the presence of the FWM effect is absent so far. 
 
%
Thus, in this manuscript, our objective is to study the linear interference between two component FLE systems in the presence of the FWM effect and explore possible localized structure and the mechanism to obtain them. 
 The present study is important because through this analysis it may be possible to obtain pulses with customized shapes.
 \\
The structure of the manuscript is the following. In the following section we introduce the model equation. The review of the methodology for obtaining the pair transition of the various localised structures is given in the third section. The results and discussion and concluding remarks will be in the fourth section.

\section{Linear interference}
\vspace{2mm}
In dimensionless form coupled FLE with FWM term:
\begin{align} 
 \label{eq:FWDU}
 iU_t & + a_1 U_{xx} +  a_2 U_{xt} +  b (|U|^2  + 2 |V|^2) U + b V^2 U^* \nonumber \\
 &+ i \rho  (|U|^2  +  |V|^2) U_x + i \rho (U V^*  + V U^*)V_x =0 \\
 i  V_t &+ a_1 V_{xx} + a_2 V_{xt} + b (|V|^2  + 2 |U|^2) V + b U^2 V^* \nonumber\\
 &+  i \rho  (|U|^2 +  |V|^2) V_x  + i \rho (U^* V + U V^*)U_x = 0  
 \label{eq:FWDV}
\end{align}

where $U$ and $V$ are complex envelopes of two wave fields. The coupled equation can describe the ultrafast pulse propagation of two mutually orthogonal optical fields in a medium where the beam is allowed to diffract along longitudinal and transverse directions. \\
Under the  gauge transformation, namely 
\begin{align}
\label{GT}
U = \sqrt{\frac{m}{|b|}}n e^{i(n x +2 mn t)} u \quad
V = \sqrt{\frac{m}{|b|}}n e^{i(n x +2 mn t)} v
\end{align}
followed by the transformation of variables, namely
\begin{align}
\label{VT}
\xi=x+m t; \quad \tau=-mn^2 t 
\end{align}
where $n=\frac{-1}{a_2} $ and $m=\frac{a_1}{a_2}$
Eqs. \ref{eq:FWDU}-\ref{eq:FWDV} are transformed into a simplified coupled equation: 
\begin{align} 
\label{eq:GFWDU}
 u_{\xi \tau}  & = u  -i 2 \sigma (|u|^2  +  |v|^2) u_\xi   -i 2 \sigma  (v^* u + v  u^*)v_\xi  \\
 v_{\xi \tau} & =v    -i 2 \sigma (|u|^2  +  |v|^2) v_\xi   -i 2\sigma  (v^* u + v  u^*)u_\xi 
\label{eq:GFWDV}  
\end{align}
   
where $\sigma=sgn(b)$. It is important to note that  eqs. \ref{eq:GFWDU} -\ref{eq:GFWDV} are the linear transformation of gauged FLE. That is taking $b>0$ the equations can be decoupled into two identical gauged FLEs, namely

\begin{align}\label{eq:FLE1}
q_{1\xi \tau}=  q_1 - 2i |q_1|^2 q_{1\xi}\\ \label{eq:FLE2}
q_{2\xi \tau}=  q_2 - 2i |q_2|^2 q_{2\xi} 
\end{align} 
such that $u =\frac{1}{2} (q_1 +q_2)$
and $v =\frac{1}{2} (q_1 - q_2)$ where $q_1$ and $q_2$ are two different independent solutions.
It is worthwhile to mention that eqs. \ref{eq:FLE1}-\ref{eq:FLE2} are the first negative hierarchy of the DNLSE. In other words, the eqs.    \ref{eq:FLE1}-\ref{eq:FLE2} are integrable even though the coupled eqs. \ref{eq:GFWDU}-\ref{eq:GFWDV} are not. 

 Let us consider the solution of eq. \ref{eq:FLE1}, a plane wave solution,
\begin{align}\label{eq:PW}
q_1 = A e^{i[k \ \xi - (\frac{1}{k} \  + 2 A^2)\tau]} 
\end{align} 
where $k$ is the wave vector and $\omega=\frac{1}{k}\ + 2 A^2$ is the frequency and $A$ is the amplitude of the plane wave. 

We consider the second solution $q_2$, a soliton solution obtained by applying a direct method, namely the Hirota bilinear method \cite{hirota2,nandy} as given in \cite{sagar}.
 \begin{align}
q_2=\frac{\alpha_1 e^{\theta+i\chi}}{\beta_1+\frac{|\alpha_1|^2}{4a^2\beta^*_1}(a+ib)^2(b+ia)e^{2\theta}}	
\end{align}
where $\theta=a(\xi+c\tau)$ and $\chi=b(\xi-c\tau)$. The parameter $c=\frac{1}{a^2+b^2}$ is the soliton velocity.\\
Earlier, linear superposition of solutions of NLSE\cite{han} and HNLSE have been reported. Here, we  investigate the solutions of the coupled FLE with FWM term using the linear superposition of component FLEs. 

\section{Soliton and plane wave pair  transition}
 
\subsection{ W-shaped, anti-dark, dark-bright soliton in plane wave background}

 
Using the linear superposition of plane wave and soliton, the analytical solutions of eqs.  \ref{eq:GFWDU} and \ref{eq:GFWDV} are given as,

\begin{align}
\label{component1}
u_1= \frac{1}{2}[A e^{i\chi_{pw}}+\frac{\alpha_1(\beta^*_1\ e^{\theta}+\beta^*_2\ e^{3\theta})}{|\beta_1|^2+(\beta_1 \beta^*_2+\beta_2 \beta^*_1)e^{2\theta}+|\beta_2|^2 e^{4\theta}}e^{i\chi_{sol}}]
\end{align} 
\begin{align}
\label{component2}
v_1= \frac{1}{2}[A e^{i\chi_{pw}} -\frac{\alpha_1(\beta^*_1\ e^{\theta}+\beta^*_2\ e^{3\theta})}{|\beta_1|^2+(\beta_1 \beta^*_2+\beta_2 \beta^*_1)e^{2\theta}+|\beta_2|^2 e^{4\theta}}e^{i\chi_{sol}}]
\end{align}
where the first term in eqs. \ref{component1} and \ref{component2} is the plane wave solution ($q_1$) of eq. \ref{eq:FLE1} having amplitude $A$  and phase   
\begin{align}
\label{phase:PW}
\chi_{pw}=k \ \xi - (\frac{1}{k} \  + 2 A^2)\tau
\end{align} 
The second term in eqs. \ref{component1} and \ref{component2} is the bright one soliton solution ($q_2$) of eq. \ref{eq:FLE2}. 

The phase of the soliton however,  is more complex due to the presence of complex coefficients, namely $\alpha_1,\ \beta_1$ and $ \beta_2$. Consequently the analysis becomes more involved. For convenience however, we identify the part of the phase function which determines soliton velocity and wave vector, namely
\begin{align}
\label{phase:SOL1}
\chi_{sol}=b\ \xi-\frac{b}{a^2+b^2}\tau 
\end{align} 
To investigate the effect of linear interference we present the solution namely $q_2$ in the following convenient form. 
\begin{align}
\label{Mix}
|u|^2= \frac{2 |A||\alpha_1| |\cosh\ z_1||\cosh\ z_2|}{\sqrt{|\beta_1||\beta_2|} \ \cosh\ (\theta+ln\frac{|\beta_2|}{|\beta_1|}+\beta_1 \beta^*_2+\beta^*_1\beta_2)}
\end{align}
where,
\begin{align}
z_1 = & X_1 + i\ Y_1;\qquad  z_2= X_2 +i Y_2& \\
X_1= & \  \theta + \frac{1}{2} ln(\frac{|\beta_1|}{|\beta_2|}) \\
Y_1= & \  \frac{\beta_1-\beta_2}{2} \\ 
X_2= & \theta + \frac{1}{2} \ ln(\frac{2|\alpha_1|\sqrt{|\beta_1||\beta_2|}|\cosh\ z_1|}{A (\cosh\ (\theta+ln\frac{|\beta_2|}{|\beta_1|}+\beta_1 \beta^*_2+\beta^*_1\beta_2)}) \\
Y_2=& \frac{(\alpha_1-\frac{\beta_1+\beta_2}{2})}{2} \qquad \qquad \qquad \qquad \qquad \text{Case-I} \\  
Y_2=& \frac{(\alpha_1-\frac{\beta_1+\beta_2}{2})}{2}+[\chi_{x(sol)}-\chi_{x(pw)}] \qquad \text{Case-II} \\  
 Y_2=& \frac{(\alpha_1-\frac{\beta_1+\beta_2}{2})}{2} + [\chi_{t(sol)}-\chi_{t(pw)}] \qquad \text{Case-III}\\
Y_2=&  \frac{(\alpha_1-\frac{\beta_1+\beta_2}{2})}{2}+[\chi_{sol}-\chi_{pw}] \qquad \qquad \text{Case-IV}
\end{align}

Notice that in eq. \ref{Mix}, the factor $ |\cosh\ z_2| $ carries the phase  information of soliton and the plane wave. The factor varies with the differences in the phases of the two superposing wave solutions. It  accounts for the modulation of the soliton into different shapes. Notice that soliton phase changes with the changes in  the soliton parameters $\alpha_1,\  \beta_1,\ a \ b $ which in turn changes $ |\cosh\ z_2| $. Hence we may obtain a variety of linear interference structures by changing the soliton parameters depending on the phase relations between soliton phase given by eq.  \ref{phase:SOL1} and plane  wave phase given by eq. \ref{phase:PW}. We narrow down the linear interference between soliton and plane wave to the following following four cases depending on their relative phases. \textit{Case-I}:\  when  the plane wave phase matches with the soliton phase. 
That is,  
\begin{align}
\label{SP}	
k&= b \\
\label{TEMP} 
\frac{1}{k} + 2 A^2 &=\frac{b}{a^2+b^2}  
\end{align}
\textit{Case-II}:\  when  $t$ component of plane wave phase matches with that of  soliton. That is, eq. \ref{TEMP} is   satisfied  but  eq.  \ref{SP} is not satisfied. \textit{Case-III}:\ when  $x$ component of plane wave phase matches with that of  soliton. That is, 
when  eq. \ref{TEMP} is not  satisfied but  eq. \ref{SP} is satisfied. \textit{Case-IV}: \ when plane wave phase does not matches with that of  soliton. None of the eqs.    
\ref{SP}, \ref{TEMP} are satisfied. \\

We further note that   from eq. \ref{TEMP} it follows that $A$ is  real when $kb\ >\ a^2 +b^2 $, imaginary when $kb\ <\ a^2 +b^2$.  In other words for
the second condition the plane wave has an additional phase of $ \frac{\pi}{2}$. Again, the phase of  soliton  depends on the soliton parameters, namely  $\alpha_1, \beta_1,a,b  $. Thus due to these additional phases we observe noticeable alteration of soliton shape profile. 

\vspace{5mm}

We represent the  different cases of soliton and plane wave   linear interference using  phase diagrams   \ref{fig:phasediagram1}-\ref{fig:phasediagram2}.   Figure  \ref{fig:phasediagram1} shows  $A$ (\text{Plane wave amplitude})   \textit{vs} $k$ (\text{wave vector}) plot drawn  with a choice of soliton parameters ($a, \ b$) such that the condition $kb\ >\ a^2 +b^2 $ in eq. \ref{TEMP} is satisfied and figure  \ref{fig:phasediagram2}  shows 
$A e^{\frac{i\pi}{2}}$  \textit{vs} $k$ plot as an example  when  $kb\ < \ a^2 + b^2$ is satisfied.   
In the figures \ref{fig:phasediagram1} - \ref{fig:phasediagram2} the blue dashed curves  represent the plot of the function \ $A=f(k)$, \ obtained from the temporal phase matching condition, namely eq. \ref{TEMP} and  the line \ $k = b (-ve)$  \ in figure  \ref{fig:phasediagram1} and \   $k = b (+ve)$ in figure  \ref{fig:phasediagram2} \ represent the spatial  phase matching condition. The interaction points of the two lines, represented by the  circled red spot  in each of the  figures \ref{fig:phasediagram1} - \ref{fig:phasediagram2}  belongs to the \textit{case-I}. All Points on the curve, $A=f(k)$ except the intersection point, belong to the \textit{case-II} and   represented by  the circled blue spots. All the points on the line \ $k= b(\pm) $ \ except the intersection points, belong to the \textit{case-III}, and  represented by the circled green spots. All other points in the $A-k$ phase space other than the curve  $A=f(k)$ and the lines $k=b(\pm)$, belong to the \textit{case-IV}, and  represented by the circled orange spot.

\begin{figure} [H]
	\centering
	\begin{tabular}{cccc}
		\includegraphics[width=0.45\textwidth]{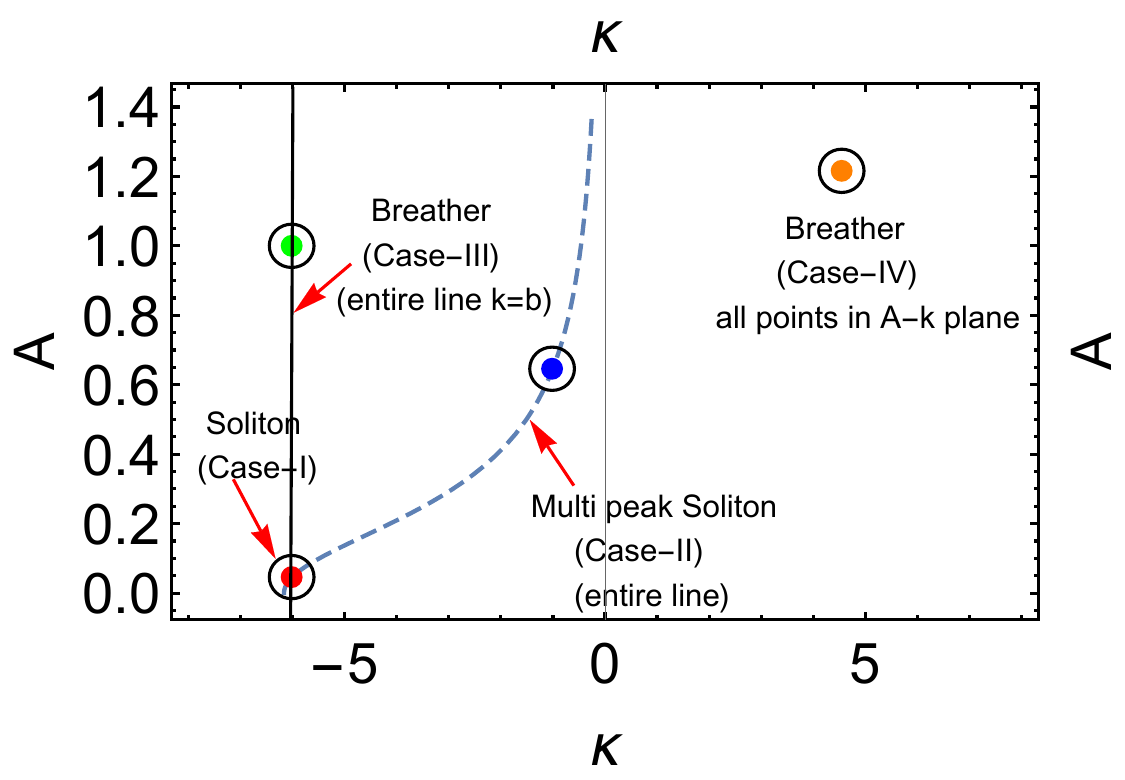} &
		\includegraphics[width=0.45\textwidth]{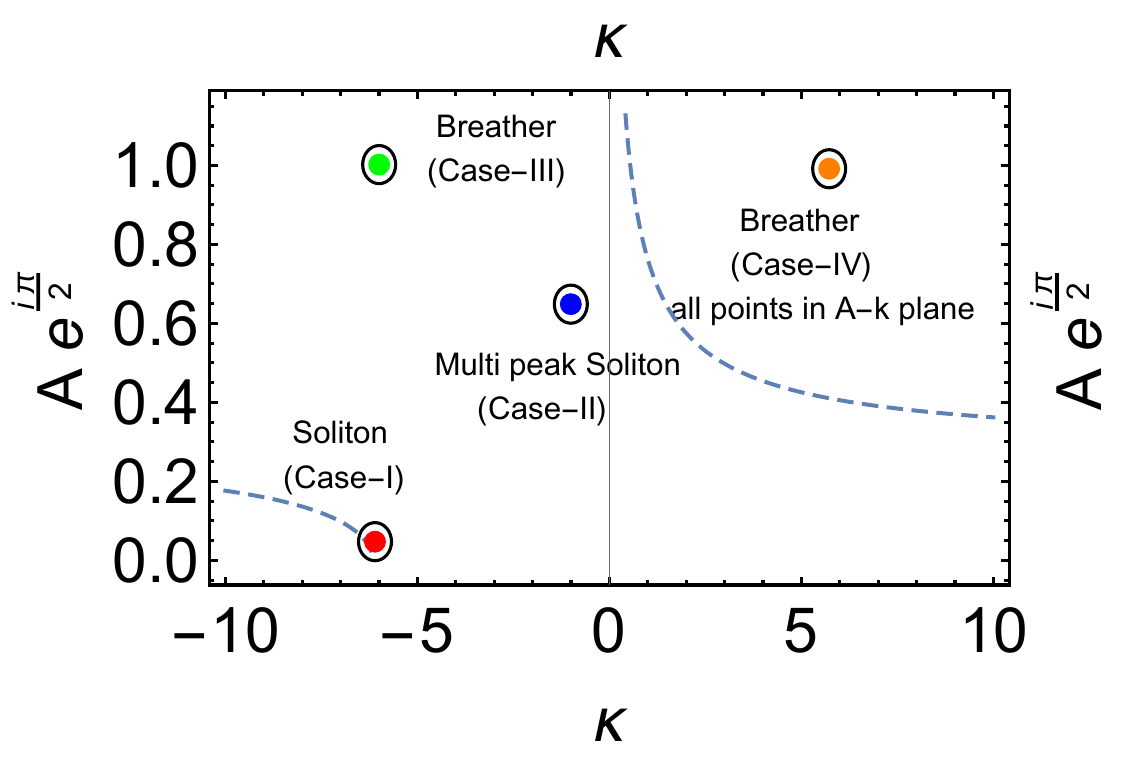} \\
		\text{(a)}  & \text{(b)}   \\[4pt]
	\end{tabular}
	\caption{Figure \ref{fig:phasediagram1} shows $A$ v/s $k$ plot drawn with  $a=-1$, $b=-6$ satisfying the condition $kb>a^2+b^2$; figure \ref{fig:phasediagram2} shows $A e^{i\frac{\pi}{2}}$ v/s $k$ plot drawn with  $a=-1$, $b=6$ satisfying the condition $kb<a^2+b^2$. Red circles corresponds to the Case-I soliton with values $a=-1$, $b=k=-6$ and $A=0.0474579$. The blue circle corresponds to values $a=-1$, $b=-6$, $k=-1$ and $A=0.647239$ representing Case-II soliton. While the green circle with parameters $a=\pm1$, $k=b=-6$ and $A=1$ is case-III breather and orange circle is a point signifying case-IV breather with parameters $a=-1$, $b=-6$, $k=4$ and $A=1.2$.}

	\label{fig:Phasediagram}
	\extralabel{fig:phasediagram1}{(a)}
	\extralabel{fig:phasediagram2}{(b)}
\end{figure}

 
\noindent
\textit{case I:} \textit{\ When  both the spatial components and temporal components of the phase are matching,  that is, $k=b$ and $\frac{1}{k}+2 A^2=\frac{b}{a^2+b^2}$}. Figure \ref{fig:case1-bright}, \ref{fig:case1-w_shaped} depicts the corresponding linear interference profiles. Notice that even when  the spatial and temporal coefficients matching conditions are met, there arise a phase difference between the plane wave and the soliton.  For instance if $b>0$ the amplitude of the plane wave has an additional  phase of $\frac{\pi}{2}$. Again, the phase of soliton changes with change in the parameters $a$ and $b$. Hence, depending on the values of the $a$ and $b$ the  plane wave and soliton may have a wide range of phase differences. Accordingly, we obtain  soliton profiles changing from ant-dark to dark to 'W'-shaped soliton. As an example an anti-dark soliton is seen in figure \ref{fig:case1-bright}.\\
  We present all the wave profiles obtained along with their conditions in the table below.   
  
\begin{center}
   \captionof{table}{Table showing conditions for various soliton profiles obtained for case-$I$.}
\begin{tabular}{|c|c|c|c|c|c|}
	\hline
	Condition &  $b>|a|$  &\multicolumn{2}{c|}{$0<b<|a|$}& $b<-|a|$ & $-|a|<b<0$\\
	\cline{3-4}
	&         & $a>0$   & $a<0$   &     &         \\
	\hline
		$\frac{1}{2}(q_1-q_2)$ & Anti-dark & Anti-dark & Dark & $W$-shaped & Dark\\
	\hline
		$\frac{1}{2}(q_1+q_2)$ & Anti-dark &  Dark   &  Anti-dark   & Anti-dark & Anti-dark \\
	\hline
\end{tabular}
\end{center}

The set of values chosen in the above table are $a=-1, b=-6$, $\alpha=1, \beta=1$. Interestingly  the profile of the soliton turns out to be  a   $W$-shaped  occur when the ratio $\frac{a}{b}< 1$.   \\

\begin{figure} [H]
	\centering
	\begin{tabular}{cccc}
		\includegraphics[width=0.3\textwidth]{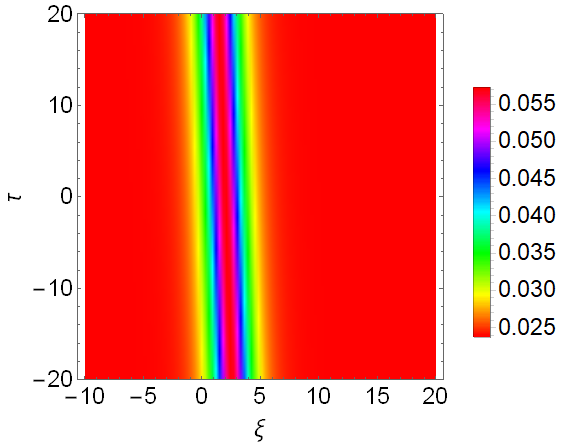} &
		\includegraphics[width=0.3\textwidth]{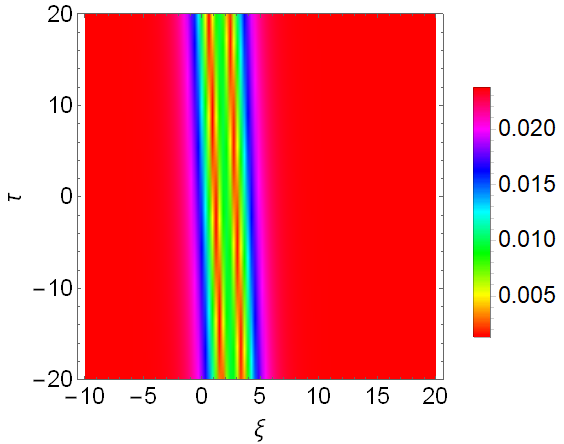} & 
		\includegraphics[width=0.3\textwidth]{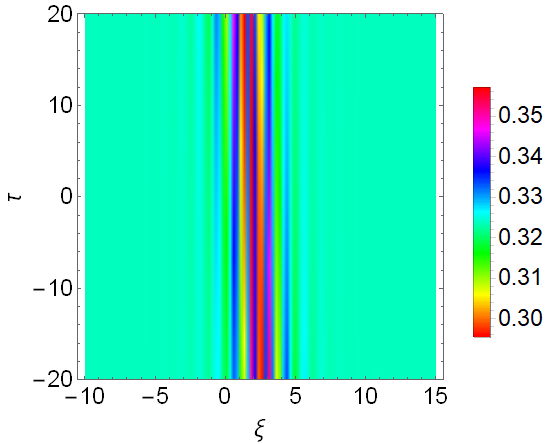} \\
		\text{(a)}  & \text{(b)} & \text{(c)}   \\[4pt]
	\end{tabular}
	\begin{tabular}{cccc}
		\includegraphics[width=0.3\textwidth]{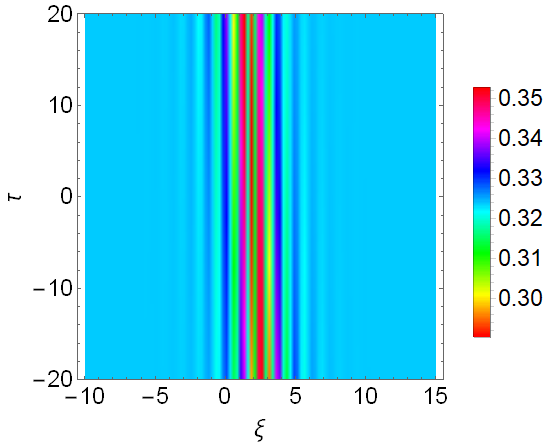} &
		\includegraphics[width=0.3\textwidth]{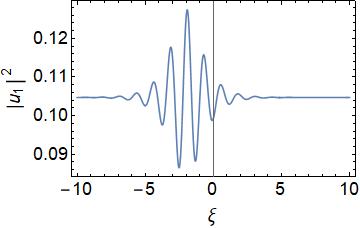} & 
		\includegraphics[width=0.3\textwidth]{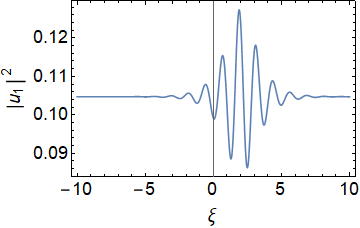} \\
		\text{(d)}  & \text{(e)} & \text{(f)}   \\[4pt]
	\end{tabular}
	\caption{The density profile of two types of soliton. (a) and
		(b) corresponds to case-$I$ soliton in components $u_1$ and $v_1$
		respectively. They correspond to the red circle in figure  \ref{fig:Phasediagram} with parameters  $a=-1$, $b=-6$, $k=-6$, $A=0.04745$. Also (c) and (d) corresponds to case-$II$ soliton, in component $u_1$ and $v_1$ respectively. They correspond to the entire black solid line in figure  \ref{fig:Phasediagram} except for the red circle. The blue circle falling on the line is a particular value of case-$II$ soliton with parameters  $a=-1$, $b=-6$, $k=-1$, $A=0.647239$. While (e) and (f) represent the shift in the position of case-$II$ soliton with the change of value in $a$. The parameters for (e) are $a=1$, $b=-6$, $k=-1$ and for (f)  $a=-1$, $b=-6$, $k=-1$. However, $\alpha_1=1$ and $\beta_1=1$ are kept constant throughout.}
	\label{fig:case1and2}
	\extralabel{fig:case1-bright}{(a)}
	\extralabel{fig:case1-w_shaped}{(b)}
	\extralabel{fig:case2-multipeak1}{(c)}
	\extralabel{fig:case2-multipeak2}{(d)} 
	\extralabel{fig:case2-multipeak_+a}{(e)}
	\extralabel{fig:case2-component_-a}{(f)}
\end{figure} 
\noindent
\\
\textit{case II:}\textit{ \ When the  temporal components are equal, but the spatial components are not matching, that is, $k\not=b$ and $\frac{1}{k}+2 A^2=\frac{b}{a^2+b^2}$}.    The corresponding linear interference profile for components $u_1$ and $v_1$ are shown in figures  \ref{fig:case2-multipeak1} and \ref{fig:case2-multipeak2} respectively. 
In the phase diagram \ref{fig:Phasediagram} this case corresponds to the case-$II$ soliton which is represented by the black solid line. Notice that due to the spatial phase  mismatch the soliton shape does not deviate much from that at  $\xi =0$, when b is positive.  This is due to vanishing background property of the soliton.
Temporally the profile remain constant. On the other hand for  $b= -ve $. The plane wave has an additional phase of $\frac{\pi}{2}$.  This results in a multipeak soliton structure which is shown in figures  \ref{fig:case2-multipeak_+a}  and \ref{fig:case2-component_-a} for component $u_{1}$. The spatial distance between the humps can be measured and is  given by $D=\frac{2\pi}{|b-k|}$. and the number of peaks is determined by the soliton's width $|\frac{1}{a}|$.    \\

\begin{figure}
	\begin{center}
		\includegraphics[width=0.5\textwidth]{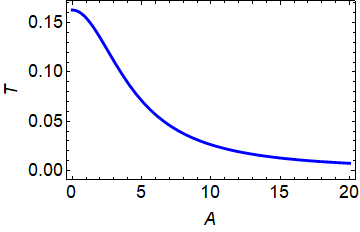}
		\caption{The figure shows variation in the time period of breathing for a case-$III$ breather with the amplitude of the plane wave. The time period decreases monotonically with the plane wave amplitude. The time period is seen to be maximum when the plane wave amplitude is nearly zero.  }
		\label{fig:timeperiod}
	\end{center}
\end{figure}

\textit{case III:}\textit{When the  temporal components are not equal, but the spatial components are matching, that is, $k=b$ and $\frac{1}{k}+2 A^2\not=\frac{b}{a^2+b^2}$}. The linear interference profile is shown in figures \ref{fig:case3-component1} and \ref{fig:case3-component2}.  In the phase diagram 
\ref{fig:Phasediagram} this case corresponds to case-$III$ breather and is represented by the straight line $k=b=-6$. The green circle represents a point on that line with parameters $a=-1$, $b=k=-6$. This localised wave having breather like structure is periodic in time and is similar to Kuznetsov-Ma(KM) breathers \cite{koznetsov}. The  hump and valley in each component are complimentary as seen in figures  \ref{fig:case3-component1} and \ref{fig:case3-component2}. The profile in this case has no significant change under the condition of inbuilt relative phase. 
Moreover, the breathing period is given by $T=\frac{2\pi}{|\frac{1}{k}+2 A^2-b\ c|}$. It is seen that the time period of breathing decreases monotonically with the increase of plane wave amplitude as shown in figure \ref{fig:timeperiod}. The time period is maximum when the amplitude of the plane wave is zero and thus induces a bright soliton.\\
\\
\textit{case IV:}\textit{When the  temporal components, as well as  spatial components, are mismatched, that is, $k\ne b$ and $\frac{1}{k}+2 A^2\ne \frac{b}{a^2+b^2}$}. This case corresponds to the case-$IV$ breather which lies in the entire domain of the figure  \ref{fig:Phasediagram} except the line and the curve of eqs. \ref{SP} and \ref{TEMP}. A typical type-IV breather is shown in the figure  \ref{fig:type3_2} with parameters $a=-1$, $b=-6$, $k=4$ and $A=1.2$.  Notice that
the condition $ b_1 \ne k$ admits a multi-peak profile while the temporal condition is responsible for the periodic nature of the wave structure.   Hence we name it as  multi-peak breather. The linear interference profile is shown in figures \ref{fig:type3_1} and \ref{fig:type3_2}. 
It is worthwhile to note that bright, dark, anti-dark, $W$-shaped profiles has been seen in \cite{Qin} for coupled NLSE with FWM term. However, in that case one has to introduce additional phase difference between plane wave and other localised waves. On the other hand, in the present case, that is in FLE with FWM the phase difference between the two linearly superposed waves  is inbuilt and depends on  $a$ and $b$, $\alpha_1$ and $\beta_1$ as well as on the sign of $b$.\\   
\begin{figure} [H]
	\centering
	\begin{tabular}{cccc}
		\includegraphics[width=0.3\textwidth]{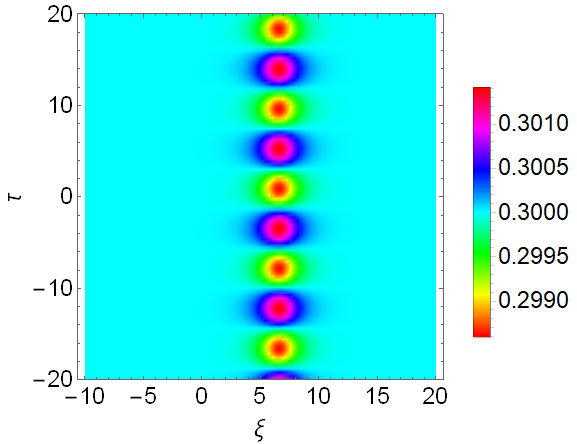} &
		\includegraphics[width=0.3\textwidth]{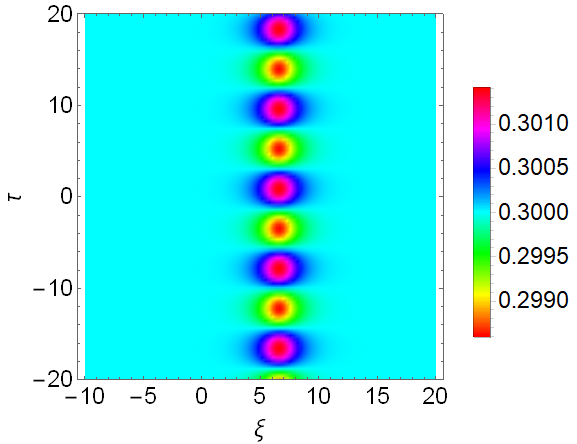} \\
		\text{(a)}  & \text{(b)}   \\[4pt]
	\end{tabular}
	\begin{tabular}{cccc}
		\includegraphics[width=0.3\textwidth]{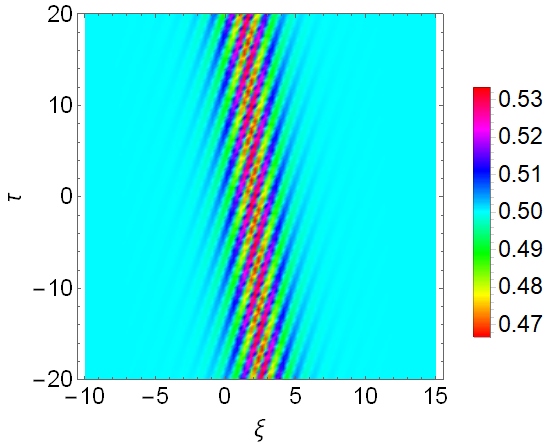} &
		\includegraphics[width=0.3\textwidth]{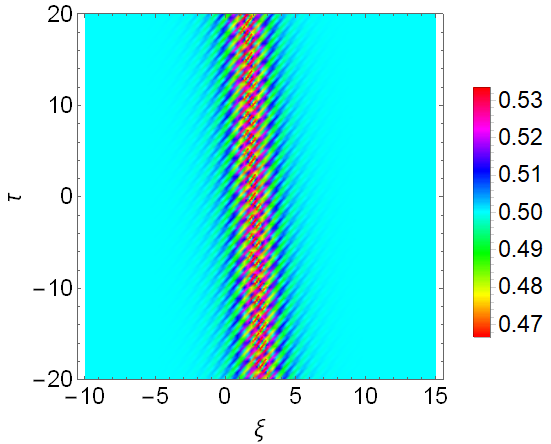} \\
		\text{(c)}  & \text{(d)}  \\[4pt]
	\end{tabular}
	\caption{The density distribution of two types of profiles where (a) and (b) corresponds to case-$III$ breather showing the exchange in the hump and valley in components $u_{1}$ and $v_{1}$ respectively.
	The parameters are $a=-1$, $b=-6$, $k=-6$, $A=1$.  While (c) and (d) are case-$IV$ breathers. The parameters are $a=-1$, $b=-6$, $k=4$, $A=1.2$. However, $\alpha_1=1$ and $\beta_1=1$ are kept constant throughout.}
	\label{fig:case3}
	\extralabel{fig:case3-component1}{(a)}
	\extralabel{fig:case3-component2}{(b)}
	\extralabel{fig:type3_1}{(c)}
	\extralabel{fig:type3_2}{(d)}
\end{figure}

We now extend the analysis to the linear interaction of plane wave and  multi-soliton. A multi soliton is the bound state of more than one solitons which are asymptotically separated.

\subsection{Plane wave and two soliton interaction}

The solution of a two soliton with a plane wave would be as follows.
\begin{align}\label{component3}
u_2=& \frac{1}{2}[A e^{i\chi_{pw}}+\\ \nonumber
&\frac{\alpha_1 e^{\theta_{1}+i \chi{1}}+ \alpha_2 e^{\theta_{2}+i \chi{2}}+\alpha_3 e^{2\theta_{1}+\theta_{2}+i \chi{2}}+\alpha_4 e^{2\theta_{2}+\theta_{1}+i \chi{1}}}{\beta_1+\beta_2 e^{2\theta_{1}}+ \beta_3 e^{2\theta_{2}}+\beta_4 e^{\theta_{1}+\theta_{2}+i(\chi_{1}-\chi_{2})}+\beta_5 e^{\theta_{1}+\theta_{2}+i(\chi_{2}-\chi_{1})}+\beta_6 e^{2(\theta_{1}+\theta_{2})}}]
\end{align}
\begin{align}\label{component4}
v_2=& \frac{1}{2}[A e^{i\chi_{pw}}-\\  \nonumber
&\frac{\alpha_1 e^{\theta_{1}+i \chi{1}}+ \alpha_2 e^{\theta_{2}+i \chi{2}}+\alpha_3 e^{2\theta_{1}+\theta_{2}+i \chi{2}}+\alpha_4 e^{2\theta_{2}+\theta_{1}+i \chi{1}}}{\beta_1+\beta_2 e^{2\theta_{1}}+ \beta_3 e^{2\theta_{2}}+\beta_4 e^{\theta_{1}+\theta_{2}+i(\chi_{1}-\chi_{2})}+\beta_5 e^{\theta_{1}+\theta_{2}+i(\chi_{2}-\chi_{1})}+\beta_6 e^{2(\theta_{1}+\theta_{2})} }]
\end{align}
where $\chi_{pw}$ is the plane wave phase as discussed earlier and the soliton parameters are $\theta_{1}=a_{1}\xi+\frac{a_{1}\tau}{a^2_{1}+b^2_{1}}$, $\theta_{2}=a_{2}\xi+\frac{a_{2}\tau}{a^2_{2}+b^2_{2}}$, $\chi_{1}=b_{1}\xi-\frac{b_{1}\tau}{a^2_{1}+b^2_{1}}$ and $\chi_{2}=b_{2}\xi-\frac{b_{1}\tau}{a^2_{1}+b^2_{1}}$

Here, one of the  two components  represent the plane wave while the other represents a two soliton wave. However, it is well known that a two soliton is the nonlinear superposition of two solitons. For convenience, we can view this as the interaction of the plane wave with two distinct  bright solitons, that is each of the solitons interacts with the plane wave following the relations namely case-$I$, case-$II$, case-$III$ and case-$IV$ as described in the previous section. It is important to  notice that, the interaction between solitons induces a phase shift in the envelope as well as the oscillatory part \cite{sagar}.\\
\begin{figure} [H]
	\centering
	\begin{tabular}{cccc}
		\includegraphics[width=0.3\textwidth]{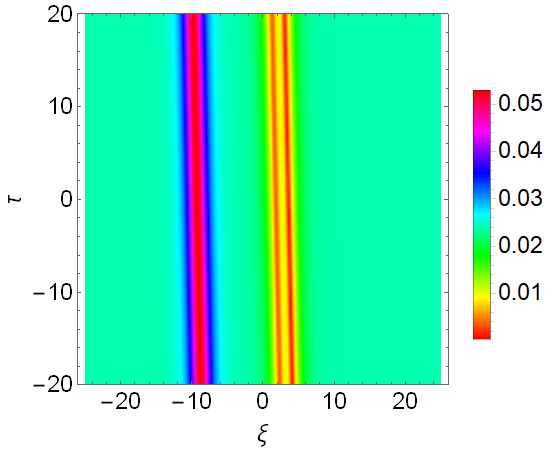} &
		\includegraphics[width=0.3\textwidth]{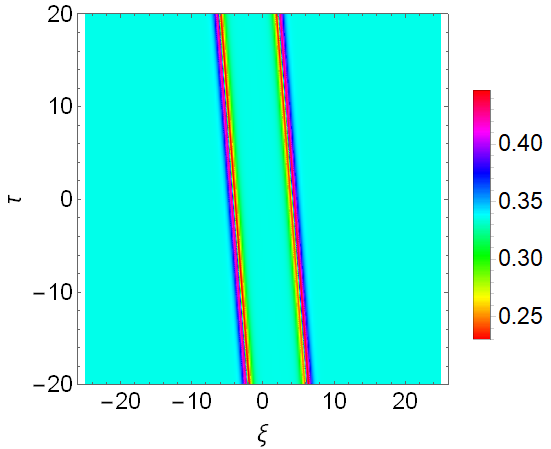} &
		\includegraphics[width=0.3\textwidth]{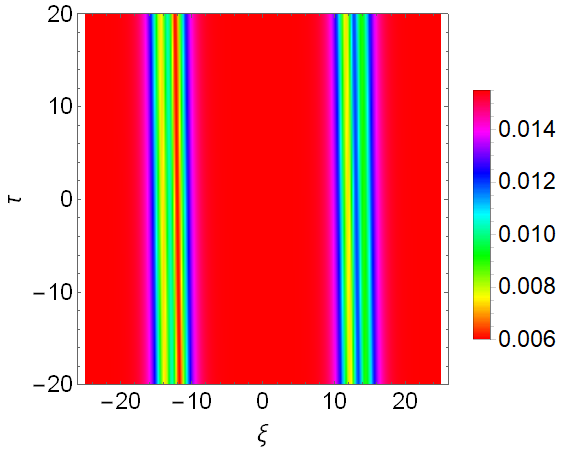} \\
		\text{(a)}  & \text{(b)} & \text{(c)}  \\[4pt]
	\end{tabular}
	\begin{tabular}{cccc}
		\includegraphics[width=0.3\textwidth]{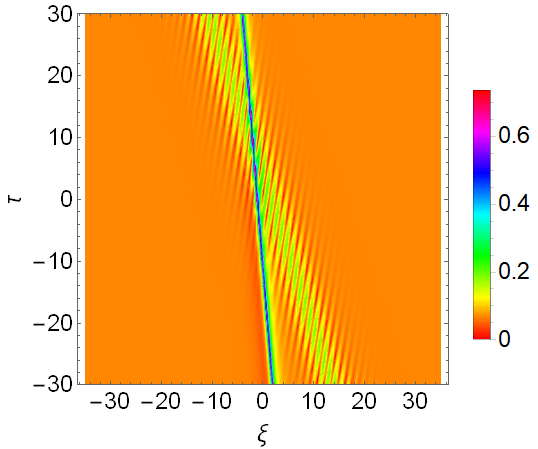} &
		\includegraphics[width=0.3\textwidth]{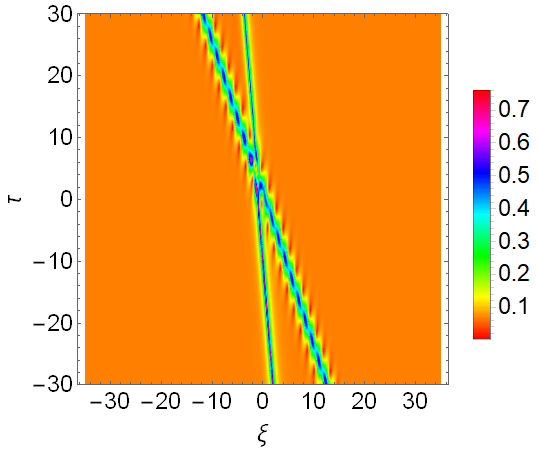} \\
		\text{(d)}  & \text{(e)}  \\[4pt]
	\end{tabular}
	\caption{The density distribution of several types of profiles where (a) and (b) corresponds to an anti-dark and $W$-shaped interaction, two dark-bright interaction in the component $u_2$ and $v_2$ respectively. While (c) shows the interaction between two $W$-shaped solitons, 
		(d)and (e) corresponds to the collision of multi-peak and anti-dark soliton and breather and anti-dark soliton respectively. The  parameters for (a) are $a_1=-1, b_1=-6, a_2=-1, b_2=-6.01$ and for (b) are $a_1=3$, $b_1=-1$, $a_2=-3$, $b_1=-1.01$. The parameters for (c) are  $a_1=1$, $b_1=-6$, $a_2=-1$, $b_1=-6.01$ while for 
		(d) the parameters are set as  $a_1=1$, $b_1=3$, $a_2=0.5$, $b_1=-1.5$ and for (e)  $a_1=1$, $b_1=3$, $a_2=1.5$, $b_1=-0.5$. Meanwhile $\alpha_1=1, \alpha_2=1, \beta_1=1$ along with  $k=b_1, A=\sqrt{\frac{b_1}{2(a^2_1+b^2_1)}-\frac{1}{2 b_1}}$ are kept constant throughout all the cases.}
	\label{fig:2ss}
	\extralabel{fig:2ss-case1-1}{(a)}
	\extralabel{fig:2ss-case1-2}{(b)}
	\extralabel{fig:2ss-case2-1}{(c)}
	\extralabel{fig:2ss-case3-1}{(d)}
	\extralabel{fig:2ss-case4-1}{(e)}
\end{figure}
 Consequently, even if both the solitons have nearly identical parameters they may not have nearly identical profiles after interaction with the plane wave that is because the second soliton has an additional phase in comparison to the phase of the first soliton. This additional phase occurs due to the nonlinear interaction of two solitons. Consequently after linear interaction with the plane wave profile of each of the solitons may considerably differ from each other. 
 \begin{center}
 	\begin{table}[htb]
 		\centering
 		\begin{tabular}{|l|c|c|c|}
 			\hline
 			
 			\multicolumn{2}{|c|}{\multirow{2}{*}{Conditions}} & \multicolumn{2}{c|}{Components} \\
 			\cline{3-4}
 			\multicolumn{1}{|l}{} & & $\frac{1}{2}(q_1-q_2)$ &  $\frac{1}{2}(q_1+q_2)$ \\
 			\hline

 			\multirow{4}{*}{\shortstack{$0<b_1<|a_1|$ \\ $0<b_2<|a_2|$}} & $a_1>0$, $a_2>0$ & \shortstack{Anti-dark+\\ Dark bright}  &  \shortstack{Dark bright+\\ Anti-dark}   \\
 			\cline{2-4}
 			&$a_1>0$, $a_2<0$ &  \shortstack{Dark bright+\\ Dark bright} &  \shortstack{Dark bright+\\ Dark bright}   \\
 			\cline{2-4}
 			&$a_1<0$, $a_2>0$ &  \shortstack{Dark bright+\\ Dark bright} & \shortstack{Dark bright+\\ Dark bright}    \\
 			\cline{2-4}
 			&$a_1<0$, $a_2<0$ &  \shortstack{Anti-dark+\\ Dark bright}  &  \shortstack{Dark bright+\\ Anti-dark}   \\
 			\hline

 			\multirow{4}{*}{\shortstack{$b_1<-|a_1|$ \\ $b_2<-|a_2|$}} & $a_1>0$, $a_2>0$ & \shortstack{W-shaped+\\ Anti-dark}  &  \shortstack{Anti-dark+\\ W-shape}   \\
 			\cline{2-4}
 			&$a_1>0$, $a_2<0$ &  \shortstack{Anti-dark+\\ Anti-dark} &  \shortstack{W-shaped+\\ W-shaped}   \\
 			\cline{2-4}
 			&$a_1<0$, $a_2>0$ &  \shortstack{Anti-dark+\\ Anti-dark} & \shortstack{W-shaped+\\ W-shaped}    \\
 			\cline{2-4}
 			&$a_1<0$, $a_2<0$ &  \shortstack{Anti-dark+\\ W-shaped}  &  \shortstack{ W-shaped+\\ Anti-dark}   \\
 			\hline

 			\multirow{4}{*}{\shortstack{$-|a_1|<b_1<0$ \\ $-|a_2|<b_2<0$}} & $a_1>0$, $a_2>0$ & \shortstack{Dark+\\ Anti-dark}  &  \shortstack{Anti-dark+\\ Dark}   \\
 			\cline{2-4}
 			&$a_1>0$, $a_2<0$ &  \shortstack{Dark bright+\\ Dark bright} &  \shortstack{Dark bright+\\ Dark bright}   \\
 			\cline{2-4}
 			&$a_1<0$, $a_2>0$ &  \shortstack{Dark bright+\\ Dark bright} & \shortstack{Dark bright+\\ Dark bright}    \\
 			\cline{2-4}
 			&$a_1<0$, $a_2<0$ &  \shortstack{Anti-dark+\\ Dark}  &  \shortstack{ Dark+\\ Anti-dark}   \\
 			\hline
 			
 			\multirow{4}{*}{\shortstack{$b_1>|a_1|$ \\ $b_2>|a_2|$}} & $a_1>0$, $a_2>0$ & \shortstack{Anti-dark+\\ Dark bright}  &  \shortstack{Dark bright+\\ Anti-dark}   \\
 			\cline{2-4}
 			&$a_1>0$, $a_2<0$ &  \shortstack{Dark bright+\\ Dark bright} &  \shortstack{Dark bright+\\ Dark bright}   \\
 			\cline{2-4}
 			&$a_1<0$, $a_2>0$ &  \shortstack{Dark bright+\\ Dark bright} & \shortstack{Dark bright+\\ Dark bright}    \\
 			\cline{2-4}
 			&$a_1<0$, $a_2<0$ &  \shortstack{Dark bright+\\ Anti-dark}  &  \shortstack{ Anti-dark+\\ Dark bright}   \\
 			\hline
 			
 		\end{tabular}
 		\caption{Conditions for various profiles obtained during the interaction of two soliton and plane wave.}
 		\label{table:2}
 	\end{table}
 \end{center}
 If the phase matching condition between the first soliton and plane wave is satisfied but not with the second soliton then the first soliton exhibits a profile as per case-$I$ and the second soliton profile is given as per case-$IV$ as discussed in one soliton. However, the time period for the second soliton may vary from a value greater than zero to a large value when $|a_1|$=$|a_2|$ and $|b_1|$ nearly equal to $|b_2|$. We find various interaction profiles such as soliton-soliton, soliton-breather, breather-breather depending on which of the four cases are satisfied with respect to first soliton. Table \ref{table:2} shows possible combinations of soliton profiles. Figure \ref{fig:2ss-case1-1}, shows  an anti-dark and $W$-shaped soliton in  $u_2$ and $v_2$ for $a_1=-1$, $b_1=-6$, $a_2=-1$, $b_1=-6.01$, 
in figure \ref{fig:2ss-case1-2}
two dark bright solitons  are shown with parameters $a_1=3$, $b_1=-1$, $a_2=-3$, $b_1=-1.01$ in component $u_2$ and  two $W$-shaped solitons in component $v_2$  \ref{fig:2ss-case2-1} with parameters $a_1=1$, $b_1=-6$, $a_2=-1$, $b_1=-6.01$.

There are however,   several other profiles are also seen for arbitrary soliton parameters such as multi-peak $\&$ anti-dark soliton, breather $\&$ anti-dark soliton interaction etc. For instance, multi-peak $\&$ anti-dark interaction is seen in figure \ref{fig:2ss-case3-1} for parameters $a_1=1$, $b_1=3$, $a_2=0.5$, $b_1=-1.5$ and breather $\&$ anti-dark interaction is seen for parameters $a_1=1$, $b_1=3$, $a_2=1.5$, $b_1=-0.5$ as shown in figure \ref{fig:2ss-case4-1}. The conditions along with the  corresponding soliton profiles are given in the table 2. Thus from the above analysis we notice that the soliton profile may be changed by changing the plane wave background. For instance a bright soliton may be transformed into dark, dark-bright or to other  desirable shape through FWM. In case of two solitons two interacting solitons may have different shape or same shape again depending upon the soliton parameters. Thus FWM allows controlling and engineering soliton profile.

\section{Conclusion and Discussion}

In this paper, we have studied the linear interference between a plane wave and bright soliton in the presence of FWM effect. We show that various localized structures, namely W-shaped
soliton, dark, anti-dark, dark-bright, breathers and multi-peak solitons  are possible through linear interaction between a plane wave and soliton. These localised structures arise due to the inbuilt relative phase between the
plane wave and bright soliton. 
The linear interference has been divided into four cases based on the spatial and temporal relation of
the coefficients of the two waves. The  linear interaction of plane wave and  two soliton interaction gives rise to several combinations of  different localized structures since nonlinear interaction between the soliton alters the soliton's phase after the interaction.
 In BEC linear interaction of waves is normal. In such interaction waves of different structures is theoretically possible; which may be worthy of investigation experimentally in future. In nonlinear optics through linear interactions several pulse structures may be generated and engineered by changing the plane wave background and soliton parameters.
 We hope that the present investigation will be useful in studying the application of FLE in nonlinear optics and various other fields.

\section*{Acknowledgement}
S Talukdar and R Dutta Acknowledge DST, Govt. of India for Inspire fellowship,  grant  nos. DST Inspire Fellowship 2020/IF200278; 2020/IF200303.

\end{document}